\documentclass[10pt,letterpaper,twocolumn]{article} %% two column, final layout

\usepackage{times,mathptmx}
\usepackage{ol2}
\usepackage{color,textcomp}
\usepackage{amsmath,amssymb}
\usepackage[english]{babel}

\begin{document}

\twocolumn[ %% activate for two-column option

%%%%%%%%%%%%%%%%%% title page information %%%%%%%%%%%%%%%%%%
\title{Coherent supercontinuum generation in a silicon photonic wire in the telecommunication wavelength range}

%\vskip-4mm

\author{Fran\c{c}ois Leo,$^{1,2,*}$ Simon-Pierre Gorza,$^3$
        St\'ephane~Coen,$^4$ Bart Kuyken,$^{1,2}$ and Gunther Roelkens$^{1,2}$}

\address{
  $^1$ Photonics Research Group, Department of Information Technology, Ghent University-IMEC, Ghent B-9000, Belgium\\
  $^2$ Center for Nano- and Biophotonics (NB-photonics), Ghent University, Belgium\\
  $^3$ OPERA-Photonique, Universit\'e Libre de Bruxelles (ULB), 50 Av. F. D. Roosevelt, CP 194/5, B-1050 Bruxelles, Belgium\\
  $^4$ Physics Department, The University of Auckland, Private Bag 92019, Auckland 1142, New Zealand\\
  $^*$ Corresponding author: francois.leo@intec.ugent.be}

\begin{abstract}%
  We demonstrate a fully coherent supercontinuum spectrum spanning 500~nm from a silicon-on-insulator photonic wire
  waveguide pumped at 1575~nm wavelength. An excellent agreement with numerical
  simulations is reported. The simulations also show that a high level of two-photon absorption can essentially
  enforce the coherence of the spectral broadening process irrespective of the pump pulse duration.
\end{abstract}


]

%%%%%%%%%%%%%%%%%%%%%%%%%%  body  %%%%%%%%%%%%%%%%%%%%%%%%%%

\noindent Supercontinuum generation is a well known phenomenon that leads to the generation of broadband light through
the nonlinear spectral broadening of narrowband sources \cite{alfano_emission_1970}. It has been studied and
demonstrated in a plethora of materials and platforms, and is now a key process for the design of broadband light
sources for many applications, such as high-precision frequency metrology \cite{jones_carrier-envelope_2000}, optical
coherence tomography \cite{hartl_ultrahigh-resolution_2001}, telecommunication \cite{smirnov_optical_2006}, or short
pulse synthesis \cite{heidt_high_2011}.

For many applications, however, bandwidth is not everything, and the coherence of the generated light is also
important \cite{nakazawa_coherence_1998, holzwarth_white-light_2001, bellini_phase-locked_2000}. In photonic crystal
fibers (PCFs), which have proved very popular for supercontinuum generation, a general criterium to obtain coherent
spectra has been developed. The coherence is highly dependent on the fiber dispersion, and on the duration and peak
power of the input pulse which can be seen as a high-order soliton \cite{dudley_coherence_2002}. A simple distinction
can be made between the ``long'' pulse regime (with input soliton order $N>16$) where noise-induced modulation
instability (MI) dominates, leading to incoherent supercontinua, and the ``short'' pulse regime (with $N<10$) where
the process is driven by initial pulse compression followed by soliton fission \cite{husakou_supercontinuum_2001},
and in which the coherence of the pump is preserved \cite{dudley_supercontinuum_2006}.

Supercontinuum generation has also recently been investigated experimentally in several chip-scale integrated
nonlinear optical platforms \cite{hsieh_supercontinuum_2007, lamont_supercontinuum_2008, halir_ultrabroadband_2012,
safioui_supercontinuum_2014, oh_supercontinuum_2014, leo_dispersive_2014}. In some of these experiments using short
pump pulses, soliton fission was experimentally demonstrated \cite{halir_ultrabroadband_2012, leo_dispersive_2014}. In particular,
soliton dynamics has been observed in silicon-on-insulator (SOI) waveguides at telecom wavelengths, suggesting that the same
mechanisms as that responsible for supercontinuum generation in PCFs can explain broadband emission in silicon waveguides
despite the inherently strong two-photon absorption (TPA) \cite{leo_dispersive_2014}. Although the occurrence of
soliton fission would suggest good coherence properties, only limited studies of the coherence of
supercontinua from integrated devices have been performed so far, based on beat note measurements \cite{phillips_supercontinuum_2011,
kuyken_octave_2014}. A complete characterization of the coherence of such a supercontinuum is however still
lacking. In this Letter, we address this point, and we demonstrate the generation of a supercontinuum with high optical
coherence across its whole spectrum in an SOI integrated waveguide pumped at telecom wavelengths.

\begin{figure}[b]
  \centerline{\includegraphics[width = 8cm]{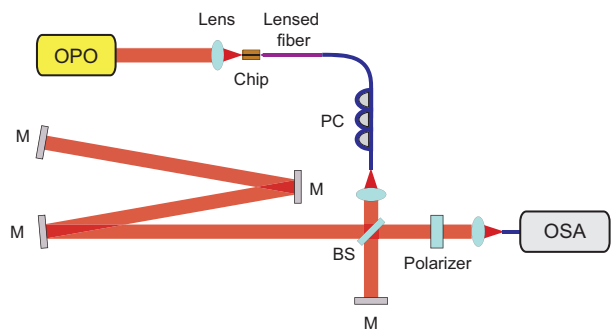}}
  \caption{Experimental setup: the SOI waveguide on a chip is pumped by an OPO. The output light is collected by a
    lensed fiber and sent to an asymmetric Michelson interferometer. The resulting interferogram is recorded on an
    optical spectrum analyzer (OSA). BS: beam splitter, PC: polarization controller, M: mirror.}
  \label{figsetup}
\end{figure}
\begin{figure}[t]
  \centerline{\includegraphics[width = 8cm]{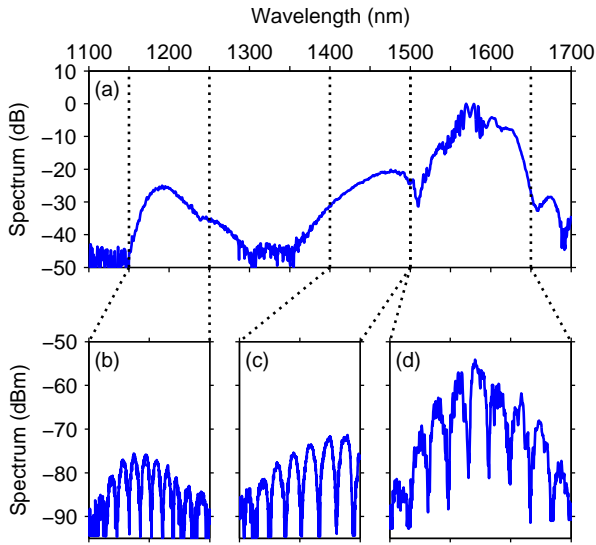}}
  \caption{(a) Experimental spectrum (1~nm spectral resolution) at the output of our SOI waveguide, normalized to a peak level of 0~dBm.
    (b)--(d) Spectral fringes measured at the output of the asymmetric Michelson interferometer with couplings optimized
    to maximize transmission in, respectively, the 1150--1250~nm, 1400--1500~nm, and 1500--1650~nm spectral windows.
    The deeply contrasted fringes indicate strong phase coherence.}
  \label{figexp}
\end{figure}

Our experiment is based on a 4~mm-long SOI waveguide with a standard height of 220~nm, a width of 700~nm, and a zero
dispersion wavelength located at 1425~nm. It is pumped with an optical parametric oscillator (OPO; Spectra Physics
OPAL) generating 150~fs pulses (full width at half maximum, FWHM) at 1575~nm wavelength with a repetition rate
$\nu_\mathrm{rep}$ of~82~MHz. Chip incoupling is performed with a microscope objective while the output light is
collected by a lensed fiber, as shown at the top of our experimental setup in Fig.~\ref{figsetup}. With a coupled
peak power of 37~W, we observe [Fig.~\ref{figexp}(a)] a broadband supercontinuum spectrum covering more than 500~nm
of bandwidth. As reported in \cite{leo_dispersive_2014}, the spectral broadening is mainly driven by soliton fission
and the subsequent emission of dispersive waves. The latter are emitted around 1200~nm wavelength, in the normal
dispersion regime \cite{akhmediev_cherenkov_1995, erkintalo_cascaded_2012}, but their precise spectral position
highly depends on the dispersion profile of the waveguide \cite{leo_dispersive_2014}.

In order to study the coherence properties of our supercontinuum spectrum, we resort to the approach pioneered by
Bellini and H\"ansch \cite{bellini_phase-locked_2000}, and which is based on the observation of spectral
interferometric fringes between two independently generated supercontinua. This method has been heavily used for
PCF-based supercontinuum generation, first numerically \cite{dudley_coherence_2002} and then experimentally
\cite{gu_experimental_2003, lu_generation_2004, turke_coherence_2007, nicholson_coherence_2008}. It provides
information about the coherence across the whole generated spectrum at once. In our experiment, we use a single SOI
waveguide as the supercontinuum source, and the two independently generated supercontinua come from subsequent pump
pulses as initially proposed for PCFs in \cite{lu_generation_2004}. Accordingly, the light collected at the output of
our waveguide is sent to an asymmetric  Michelson interferometer (see Fig.~\ref{figsetup}) with a delay $\tau =
1/\nu_\mathrm{rep} + \delta t$ close to the time period between two subsequent pump pulses. The extra delay $\delta
t$ is chosen to have a reasonable density of spectral fringes, but not so high as to be limited by the resolution of
the optical spectrum analyzer (OSA). The visibility of the spectral fringes recorded on the OSA was maximized by
introducing a polarizer at the output of the interferometer.

Our coherence measurements have been taken separately over three spectral windows (1150--1250~nm, 1400--1500~nm, and
1500--1650~nm) because light is delivered and collected into and out of our bulk interferometer through optical
fibers, and the coupling efficiencies depend on wavelength. Figs.~\ref{figexp}(b)--(d) show spectral interferograms
in these three spectral windows, in each case measured after re-optimizing the couplings. The vertical axis in
Figs.~\ref{figexp}(b)--(d) directly reflects the actual power density at the output of the interferometer, showing
how the OSA noise floor (at $-95$~dBm) limits our measurements. Spectral fringes spaced by about 10--15~nm are
clearly visible in all three windows. The fringe contrast is excellent throughout the spectrum, typically exceeding 20~dB,
indicating excellent phase coherence with the pump across the span of our supercontinuum. In particular, we note that
the dispersive-wave peak around 1200~nm wavelength is coherent too, despite being effectively separated from the rest
of the spectrum.

\begin{figure}[t]
  \centering
  \includegraphics[width = 8cm]{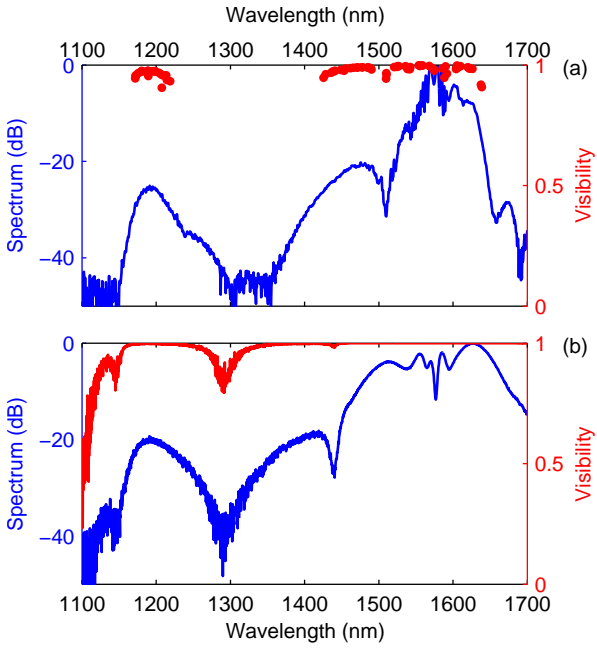}
  \caption{Comparison of (a) experimental and (b) numerical results. Blue: Spectrum at the output of the
    waveguide. Red: Visibility of spectral fringes, in (a) as observed at the output of our asymmetric
    Michelson interferometer, and in (b) obtained with stochastic numerical simulations.}
  \label{figsim}
\end{figure}

The coherence of the generated light can be quantified by extracting the fringe visibility as a function of
wavelength from our measurements, $V = (I_\mathrm{max}-I_\mathrm{min})/(I_\mathrm{max}+I_\mathrm{min})$, where
$I_\mathrm{max}$ and $I_\mathrm{min}$ correspond to the maxima and minima of each of the fringes. The results of these
calculations are plotted as red dots in Fig.~\ref{figsim}(a), where the corresponding wavelength has been set for
each point halfway between the maximum and the minimum of the fringe. Note that the high density of points in that
figure results from taking measurements from four independently recorded interferograms in each spectral window. For
comparison, Fig.~\ref{figsim}(a) also shows the overall generated supercontinuum spectrum [same as in
Fig.~\ref{figexp}(a)]. We can see that the coherence is very high across the whole spectrum, with an average
visibility of 97\,\% which appears to be limited only by the noise floor of the OSA. Our analysis therefore proves
that broad coherent supercontinua can be generated in a silicon photonic wire pumped at telecom wavelengths. We must note that the
input soliton order is relatively large in our experiment, $N \simeq 20$. Supercontinua generatd in PCF's are typically incoherent
for such large values of $N$ \cite{dudley_supercontinuum_2006}. Here, we believe that coherence can be obtained for
larger $N$ because of the large TPA of silicon, which effectively limits the number of solitons that can exist after
fission as described in \cite{leo_dispersive_2014}.

To understand our results better, we have compared our experimental measurements with stochastic numerical
simulations. Our model is based on a nonlinear Schr\"odinger equation with a complex nonlinear parameter accounting
for TPA [$\gamma = (234 + 44i)\ \mathrm{W^{-1} m^{-1}}$], and also includes free carrier absorption (FCA) and free
carrier dispersion (FCD). That model was shown in \cite{leo_dispersive_2014} as appropriate for supercontinuum
generation in silicon with short pump pulses. Note that apart from the parameters explicitly listed here, we have used the
same parameter values as in that earlier work. The degree of first-order coherence $g_{12}^{(1)}$ of the generated
light can be evaluated from the classical formula \cite{dudley_coherence_2002, gu_experimental_2003}
\begin{equation}
  g_{12}^{(1)}(\lambda) = \frac{\left<E_1^*(\lambda)E_2(\lambda)\right>}%
                               {\sqrt{\left<|E_1(\lambda)|^2\right>\left<|E_2(\lambda)|^2\right>}}\,,
  \label{Eqg12}
\end{equation}
where the angle brackets denote ensemble averages over supercontinuum spectra $E_{1,2}(\lambda)$ obtained from
separate simulations with different realizations of input noise. In practice, input noise is taken into consideration
by adding one photon per mode with random phase to the initial condition. We have performed 20~independent
simulations, and the averages were calculated over each pair of calculated output fields. The visibility of spectral
fringes can then be obtained as
\begin{equation}
  V(\lambda)=\frac{2\sqrt{I_1(\lambda)I_2(\lambda)}}{I_1(\lambda)+I_2(\lambda)} \left| g_{12}^{(1)}(\lambda) \right|
  \label{Eqvis}
\end{equation}
where $I_{1,2}$ correspond to the averaged spectral intensities of each arm of the interferometer. Note that the
fringe visibility constitutes a lower bound of the modulus of the coherence function, and that they are equal only if
the powers retrieved from both arms of the interferometer are balanced, which was nearly the case in our experiment.

Our numerical simulation results are presented in Fig.~\ref{figsim}(b), in which we show both the average spectral
intensity at the waveguide output (blue) and the visibility of spectral fringes (red). Both curves are in excellent
agreement with the experimental results [Fig.~\ref{figsim}(a)], confirming that spectral broadening in our conditions
is independent from input noise. We believe this is the first clear demonstration of supercontinuum coherence in the
presence of carrier dynamics. Carriers are well known to highly affect the nonlinear dynamics of silicon wires pumped
at telecommunication wavelengths in the case of longer pulses, through both FCD and FCA \cite{yin_impact_2007}. As
the effect of carriers is much weaker in the femtosecond regime \cite{yin_soliton_2007}, our results allow to
envision the use of silicon waveguides to extend frequency combs generated for example by Erbium doped fiber lasers.

\begin{figure}[b]
  \centering
  \includegraphics[width = 8cm]{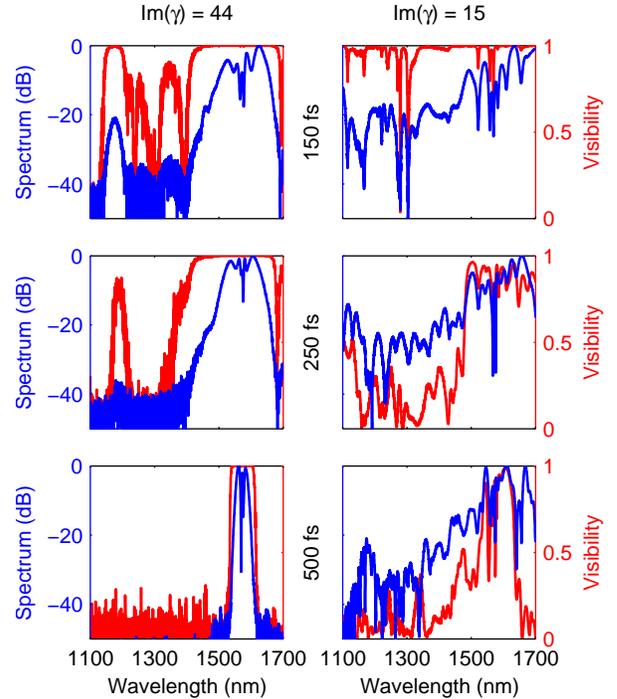}
  \caption{Simulated supercontinuum spectra (blue) and their coherence (visibility, red) for increasing pump pulse
    durations (from top to bottom, $T_0=150$~fs, 250~fs, and 500~fs) and for two different TPA strengths
    [$\mathrm{Im}(\gamma)=44$ and $15\ \mathrm{W^{-1}m^{-1}}$ in the left and right column, respectively]. Waveguide
    length: 1~cm; Pump peak power: 20~W.}
  \label{figextrasim}
\end{figure}
To understand further the influence of the nonlinear losses and the pump pulse duration on the coherence of
supercontinuum generation in silicon, we have performed additional numerical simulations whose results are summarized
in Fig.~\ref{figextrasim}. Here we used 20~W peak power pump pulses and a 1~cm-long waveguide, slightly longer than
in our experiment, but otherwise with the same width, thickness, and dispersion parameters as before. The pump pulse
duration (FWHM) increases from 150~fs (top, as in our experiment) to 250~fs (middle), and 500~fs (bottom). In the
left column, we have used the same nonlinear parameter $\gamma$ as in Fig.~\ref{figsim}, and that corresponds to our
experimental conditions. In contrast, results in the right column have been obtained for a reduced TPA, $\gamma =
(234 + 15i)\ \mathrm{W^{-1} m^{-1}}$, corresponding to a better figure of merit like that found at longer pump
wavelengths or in other materials like amorphous silicon \cite{safioui_supercontinuum_2014, liu_bridging_2012}.

The results shown in the top-left of Fig.~\ref{figextrasim}, with 150~fs pump pulses and high TPA, are
similar to our current experiment, i.e., a broad supercontinuum with perfect coherence across the whole bandwidth. We
note that the coherence is maintained here despite the longer waveguide considered (1~cm versus 4~mm) and it
indicates that the supercontinuum spectrum observed in \cite{leo_dispersive_2014} was also likely fully coherent. As
the pump pulse duration increases (left of Fig.~\ref{figextrasim}, middle and bottom graphs), the spectral width of
the supercontinuum shrinks, but an excellent coherence is nevertheless maintained across the generated bandwidth.
This is in stark contrast with supercontinuum generation in PCF where the coherence is observed to collapse for long
pulse durations \cite{dudley_supercontinuum_2006}. As the TPA is reduced (right column), the supercontinuum appears
broader for all pump pulse duration considered, which is a simple consequence of the lower losses enabling higher
power throughout the waveguide. However, in this case, the coherence does severely degrade for the longer pump
pulses, with almost no coherence preserved with 500~fs pump pulses.

Overall, TPA seems to enforce the coherence of the supercontinuum generation process. If a spectrum broad enough for
one's application can be obtained despite the strong nonlinear losses, it will be coherent, irrespective of pump
pulse duration or input soliton order. Our experiment [Fig.~\ref{figsim}(a)] hits that sweet spot. However, the
spectral broadening quickly saturates with the peak power, and using longer pump pulses lead to a collapse of the
spectral width. We can relate these observations to the increase of the fission length for longer pump pulse
durations. As the fission is delayed, other mechanisms take over the spectral broadening. In PCF, or waveguides with weak TPA,
Modulation instability (MI) kicks in but, being noise driven, leads to poor coherence. High TPA quenches MI, leaving only self-phase
modulation, which is a coherent process, but then spectral broadening is limited by the fast nonlinear drop of peak
power.

In conclusion, we have studied experimentally and numerically the coherence properties of supercontinuum generation
in a silicon wire waveguide pumped at telecommunication wavelengths where the two-photon based
nonlinear losses are significant. We have demonstrated a fully coherent 500~nm-wide supercontinuum spectrum, spanning
from 1200~nm to 1700~nm. Numerical simulations agree very well with experimental observations and explain the role
of TPA. Coherence appears to be preserved because the supercontinuum mainly arises through soliton fission despite
the high value of the input soliton order, which is in practice reduced by TPA. Our results also reveal that in our
experimental conditions, with relatively short 150~fs pump pulses, free carriers do not significantly affect the
spectral broadening nor the coherence of the process.

This work is supported by the Belgian Science Policy Office (BELSPO) Interuniversity Attraction Pole (IAP) programme
under grant no.~IAP-6/10 and by the FP7-ERC-MIRACLE project. The participation of S.~Coen to this project was made
possible thanks to a Research~\& Study Leave granted by The University of Auckland and to a visiting fellowship from
the FNRS (Belgium). Bart Kuyken acknowledges the special research fund of Ghent University (BOF) for a fellowship. The authors would like to thank Laurent Olislager for fruitful discussions.
%\bibliographystyle{osa}      %% With titles
%\bibliographystyle{optlet}   %% Without titles
%\bibliography{sc,sc-bulk,sc-coh,sc-silicon}

\clearpage

\def\refname{References with titles}

\end{document}